\documentclass[aps,pra,reprint,amssymb,amsmath, superscriptaddress,noeprint]{revtex4-2} 

\usepackage{graphicx}
\usepackage{dcolumn}
\usepackage{bm}
\usepackage[colorlinks=true,citecolor=blue,linkcolor=blue,urlcolor=blue]{hyperref}
\usepackage{orcidlink}
\usepackage{physics}
\newcommand{\dagg}{^{\dag}}

\begin{document}

\title{Amorphous quantum magnets in a two-dimensional Rydberg atom array} 

\author{Sergi Julià-Farré\,\orcidlink{0000-0003-4034-5786}}
\let\comma,
\thanks{These two authors contributed equally to this work.}
\affiliation{PASQAL SAS, 7 rue Léonard de Vinci - 91300 Massy, Paris, France}

\author{Joseph Vovrosh\,\orcidlink{0000-0002-1799-2830}}
\let\comma,
\thanks{These two authors contributed equally to this work.}
\affiliation{PASQAL SAS, 7 rue Léonard de Vinci - 91300 Massy, Paris, France}

\author{Alexandre Dauphin\,\orcidlink{0000-0003-4996-2561}}
\email{alexandre.dauphin@pasqal.com}
\let\comma,
\affiliation{PASQAL SAS, 7 rue Léonard de Vinci - 91300 Massy, Paris, France}

\begin{abstract}
Amorphous solids, i.e., systems which feature well-defined short-range properties but lack long-range order, constitute an important research topic in condensed matter. While their microscopic structure is known to differ from their crystalline counterpart, there are still many open questions concerning the emergent collective behavior in amorphous materials. This is particularly the case in the quantum regime, where the numerical simulations are extremely challenging. In this article, we instead propose to explore amorphous quantum magnets with an analog quantum simulator. To this end, we first present an algorithm to generate amorphous quantum magnets, suitable for Rydberg simulators of the Ising model. Subsequently, we use semiclassical approaches to get a preliminary insight of the physics of the model. In particular, for ferromagnetic interactions, we calculate mean-field phase diagrams, and use the linear-spin-wave theory to study localization properties and dynamical structure factors of the excitations. For antiferromagnetic interactions, we show that amorphous magnets exhibit a complex classical energy landscape by means of simulated annealing. Finally, we outline an experimental proposal based on Rydberg atoms in programmable tweezer arrays, thus opening the road towards the study of amorphous quantum magnets in regimes difficult to simulate classically.
\end{abstract}

\maketitle

\section{Introduction}\label{sec:intro}
\begin{figure}[t]
    \centering
    \includegraphics[width=\columnwidth]{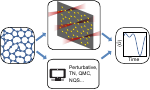}
    \caption{A schematic showing the process to perform simulations of an amorphous solid either through classical methods like e.g., perturbation theory, Tensor Networks (TN), Quantum Monte Carlo (QMC) and Neural Quantum States (NQS), or through the use of Rydberg atoms in optical tweezers.}
    \label{fig:frontpanel}
\end{figure}

The study of many-body interacting quantum systems is arguably one of the major challenges of modern physics~\cite{thouless_quantum_1972,coleman_introduction_2015}. From a classical simulation perspective, extracting their relevant physics is a hard problem due to the exponential increase of the Hilbert space with the system size. This fundamental bottleneck is typically circumvented in state-of-the-art numerical methods~\cite{fehske_exact_2008,verstraete_matrix_2008,orus_practical_2014,schmitt_quantum_2020} by taking advantage of the special properties of physically relevant states~\cite{hastings_area_2007}, as well as intrinsic system symmetries, e.g., discrete translational symmetries in crystals. In this context, quantum simulators~\cite{feynman_simulating_1982} offer a promising alternative~\cite{altman_quantum_2021,trabesinger_quantum_2012,georgescu_quantum_2014} to their classical counterpart. These versatile quantum systems, used to mimic complex quantum systems in a controllable way, allow for the exploration of complex regimes that are hardly accessible with classical numerics. A particularly relevant example is the case of 2D systems lacking intrinsic spatial symmetries, such as disordered lattices, quasicrystals and amorphous solids.  While the quantum simulation of disordered systems~\cite{menu_anomalous_2020} and quasicrystals~\cite{mace_quantum_2016,schreiber_observation_2015,sbroscia_observing_2020,gottlob_hubbard_2023,choi_exploring_2016,bordia_probing_2017,yu_observing_2023} has been recently discussed, in this work we propose to simulate amorphous quantum magnets with Rydberg atoms~\cite{saffman_quantum_2010} trapped in optical tweezers~\cite{browaeys_many-body_2020,henriet_quantum_2020}, as sketched in Fig.~\ref{fig:frontpanel}, and benchmark the properties of these systems in the quantum regime with a perturbative approach.

Since the seminal works on non-crystalline materials~\cite{zachariasen_atomic_1932,weaire_structure_1976,mott_conduction_1969,n_f_mott_electrons_1977}, a lot of theoretical~\cite{zallen_physics_1998,thapa_ab_2023, stachurski_structure_2011,petrakovskii_amorphous_1981, du_atomistic_2022,christie_review_2023} and experimental~\cite{moss_evidence_1969,mcmillan_electron_1981,yang_determining_2021,corbae_observation_2023} efforts have been put to determine the atomic structure and electronic properties of amorphous materials. In an amorphous solid, the lack of a crystal translational symmetry is also accompanied by the absence of long-range orientational order, in contrast to quasicrystals. 
Still, amorphous solids differ from fully disordered systems in that they exhibit well defined short-range properties. 
In particular, their atomic constituents have preferred bond lengths and bond angles, giving rise to a well defined coordination number $C$, i.e. the average number of nearest-neighbors (NN) per atom. Remarkably, these unique properties of amorphous solids have already found technological applications such as, for instance, the use of amorphous silicon to make solar cells and thin-film transistors \cite{street_technology_1999}. 

Amorphous solids are also interesting in the context of emergent many-body phases. While this field has historically focused on crystalline (lattice) systems, a few works have already considered classical phase transitions in amorphous magnets~\cite{aharony_critical_1975,blote_cluster_2002,kawamura_simplified_1983,kawamura_simplified_1985}, and there is also a growing interest in the study of quantum many-body phases in amorphous materials. The latter has led to the concepts of amorphous superconductors~\cite{bergmann_amorphous_1976, tsuei_amorphous_1981, manna_noncrystalline_2022}, amorphous topological insulators~\cite{agarwala_higher-order_2020,kim_fractionalization_2023,corbae_amorphous_2023}, and amorphous spin liquids~\cite{cassella_exact_2023,grushin_amorphous_2023}. However, due to the numerical complexity of simulating quantum amorphous solids, these works were restricted to single-particle ~\cite{bergmann_amorphous_1976,manna_noncrystalline_2022,agarwala_higher-order_2020}, mean-field~\cite{kim_fractionalization_2023} approximations, integrable models~\cite{cassella_exact_2023}, or exact diagonalization studies of small systems~\cite{grushin_amorphous_2023}. Analog quantum simulators are thus a good candidate to further understand the collective phenomena of quantum amorphous solids from a microscopic perspective. 

In this direction, in this article we present a toolbox to investigate amorphous magnets described by quantum Ising Hamiltonians that are suitable for simulators based on neutral atoms trapped with optical tweezers~\cite{saffman_quantum_2010,browaeys_many-body_2020,henriet_quantum_2020}. This platform stands out to investigate amorphous magnets, since both the Ising~\cite{scholl_quantum_2021} and XY~\cite{de_leseleuc_optical_2017} models can be synthesized by taking advantage of large-dipole Rydberg transitions in setups that can be scaled to hundreds of qubits. Furthermore, in contrast to digital devices with limited connectivity~\cite{barredo_synthetic_2018, henriet_quantum_2020}, or ultracold atoms trapped in optical lattices~\cite{lewenstein_ultracold_2012,bloch_many-body_2008}, the freedom in the atomic register layout offered by programmable tweezer arrays allows one to directly consider an amorphous configuration without the need for additional resources. 

With this quantum simulation goal in mind, we begin by proposing a classical variational method that generates samples of amorphous layouts from a few physical inputs such as the system size, and the average and variance of the coordination number. Importantly, this protocol takes into account a power-law dependence of interactions with the interatomic distance, and thus allows one to build physical real-space amorphous Hamiltonians. In view that solving accurately these Hamiltonians represents a hard numerical challenge, we make a first step into understanding their underlying physics by using a mean-field ansatz for ground states and the linear spin wave theory (LSWT) for the excitation spectrum. This allows us to simulate large system sizes for given parameter regions in which quantum effects can be treated as perturbations to the classical solution. In particular, we study both groundstate ferromagnetic transitions, as well as localization properties and dynamical structure factors of LSWT excitations. In the case of antiferromagnetic interactions, due to the complexity of the quantum regime, we limit ourselves to study the glassy classical energy landscape of amorphous systems by using simulated annealing (SA).
In order to go beyond this preliminary numerical study, we also describe a concrete experimental proposal to investigate both the general equilibrium and non-equilibrium properties of amorphous magnets, both in the ferromagnetic and antiferromagnetic regimes, by means of programmable tweezer arrays of Rydberg atoms.

This article is structured as follows. In Section~\ref{sec:model}, we describe the Hamiltonian model considered throughout this article. In Section~\ref{sec:generation}, we present a variational protocol to generate amorphous quantum Ising models. Section~\ref{sec:lswt} consists on a brief review of the LSWT, which is used to diagnose the physics of amorphous magnets in Sections~\ref{sec_mf_pd}-\ref{sec_dsf}. More specifically, in Section~\ref{sec_mf_pd} we compute their mean field phase diagram, while in Sections~\ref{sec_loc} and \ref{sec_dsf} we study, respectively, the localization properties and dynamical structure factors of their LSWT spectrum. In Section~\ref{sec:simulated_annealing}, we discuss the complexity of the antiferromagnetic energy landscape with classical SA. Section~\ref{sec:experiment}  is devoted to the experimental proposal for realizing the present model with neutral atom simulators. Finally, we draw the conclusions in Section~\ref{sec:conclusion}, where we also discuss possible future directions for research on amorphous magnets.

\section{\label{sec:model}Model}

\begin{figure*}
    \centering
    \includegraphics[width=\textwidth]{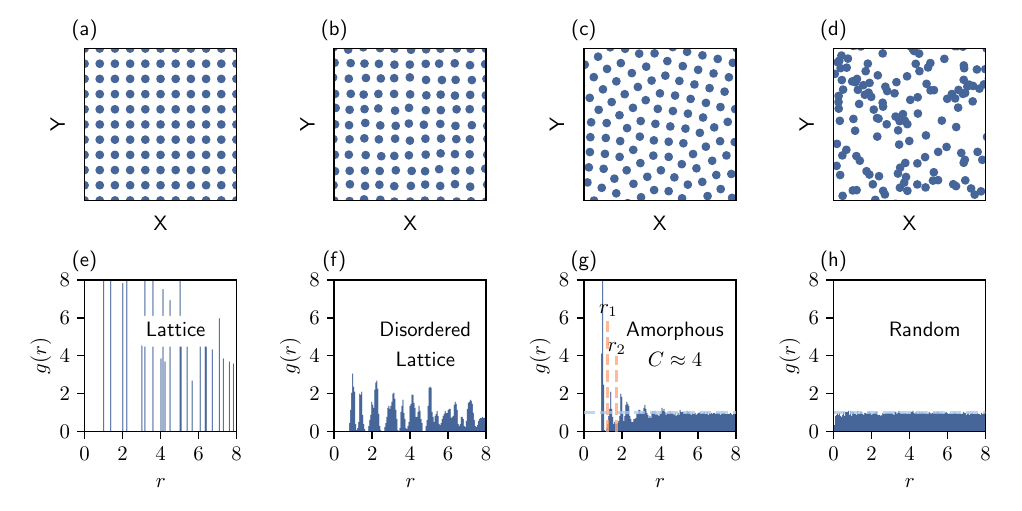}
    \caption{The radial distribution function for a square lattice, disordered square lattice, amorphous solid and a random configuration. Panels (a)-(d) show the atomic positions of a given realisation of each of these systems. Panels (e)-(h) show the corresponding radial distribution functions, $g(r)$. In panel (e) and (f) we see both well defined short and long-range order that is a consequence of the lattice and underlying lattice of the square and disordered square lattices respectively. Both panel (g) and (f) show no long-range order, i.e. $g(r)\rightarrow1$ for large $r$, note, we have highlighted this via a horizontal dashed line at $g(r)=1$. However, panel (g), which corresponds to the amorphous solid, shows clear peaks at small $r$. This reflects the short-range order in this two dimensional system. The first and second minima of $g(r)$ are highlighted as $r_1$ and $r_2$. These can be used to define both NN atomic pairs and NNN atomic pairs in the amorphous solid.}
    \label{fig:lattice_to_amorphous}
\end{figure*}

In this article we consider the transverse field Ising model (TFIM) describing $N$ interacting spin $S=\frac{1}{2}$ sites in two spatial dimensions 
\begin{equation}\label{eq:ising_hamiltonian}
H=-\sum_{i<j}J_{ij}S^z_iS^z_j+h_x\sum_i S^x_i,
\end{equation}
where $S^\alpha_i=\frac{1}{2}\sigma^\alpha_i$, in which $\sigma^\alpha_i$ are the usual Pauli matrix operators on $i$-th site located at a position $\mathbf{r}_i$. 
Here $h_x\geq 0$ is a spatially uniform transverse field, and the spin-spin interaction, $J_{ij}$, is assumed to follow a power-law decay with distance, i.e., $J_{ij}=J_0/|\mathbf{r}_i-\mathbf{r}_j|^\alpha$.
Throughout this work we will use $\alpha=6$ to reflect the Ising model realisable in a neutral atom quantum simulator \cite{henriet_quantum_2020}.

The magnetic couplings of the Hamiltonian in Eq.~\eqref{eq:ising_hamiltonian} are therefore fully defined through the positions $\mathbf{r}_i$ of the spins, which we consider to correspond to a two-dimensional amorphous solid. As previously mentioned, these systems lack long-range spatial order, but its short-range order leads to a well defined $C$, i.e., the average number of NN for each atom~\cite{zallen_physics_1998}. 

However, to define which atoms are NN in an amorphous solid is more involved than in a lattice system where the separation of every $n^{\text{th}}$ order NN is fixed.

To generalise the notion of each $n^{\text{th}}$ order NN we use the radial distribution function, $g(r)$.
This function describes how the density of the solid varies as a function of distance, $r$, from a given qubit. 
In particular, the radial distribution function for a material is given by
\begin{equation}
    g(r) = \frac{1}{\rho}\langle\sum_i\delta(r-r_i))\rangle,
\end{equation}
where $\rho$ is the density of the solid prepared, $r_i$ is the distance between the $i^{\text{th}}$ atom and the reference atom and the summation is taken over all atoms other than the reference atom. However, in practice a representative subset of atoms is typically used. 
For a given $g(r)$, the NN coordination number of a solid is given by
\begin{equation}
    n_1 = 2\pi\int_0^{R_1}rg(r)dr,
\end{equation}
where $R_1$ is the first minima in the radial distribution function.
Two atoms are considered $n^{\text{th}}$ neighbors if their separation, $r$, is such that $R_{i-1}<r<R_i$ in which $R_0=0$. 
Note, while in systems with long-range order we are able to define $n^{\text{th}}$ order NN for large $n$, in an amorphous solid we are limited to small only a few orders of nearest neighbours. 

In Fig.~\ref{fig:lattice_to_amorphous} we show the radial distribution function for a square lattice, disordered square lattice, an amorphous solid and a random configuration. 
The well defined lattice structure of the square lattice manifests itself as sharp peaks at fixed values of $r$, see Fig.~\ref{fig:lattice_to_amorphous}(e). 
The radial distribution function for the disordered square lattice shares this structure but the added positional disorder leads to a spreading of each peak, see Fig.~\ref{fig:lattice_to_amorphous}(f). 
Despite this spreading, the peaks are clearly visible even at large $r$ - this is a signature of the inherent long-range order of the system due to the underlying lattice. 
For an amorphous solid it is known that the short-range order manifests as well-defined peaks in $g(r)$ for small $r$, whereas $g(r)$ tends to one at large $r$ due to the lack of long-range order, see Fig.~\ref{fig:lattice_to_amorphous}(g). 
This trend to $g(r)=1$ at large atomic separation reflects complete randomness that we exemplify by plotting the radial distribution function for a random configuration in Fig.~\ref{fig:lattice_to_amorphous}(h).

Note, in the remainder of this article we standardise the interaction landscape by setting both $J_0$ and the minimum distance between two atoms, $\min_{i,j}{|\mathbf{r}_i-\mathbf{r}_j|}$, to unity. 
As a result, the interaction strengths in this model are such that $J_{ij} \leq 1$. 
Therefore, to allow for a direct comparison with known results in lattices we define $\bar{J}$ as the average NN interaction and scale our transverse field strength with respect to this quantity, i.e., we will consider $h_x/\bar{J}$.

\section{\label{sec:generation}Amorphous Solid Generation}

While some works have had success approximating amorphous solids through completely random atomic positions~\cite{agarwala_higher-order_2020, manna_noncrystalline_2022}, in order to accurately simulate the physics of these materials in our model both the known short-range order and lack of long-range order need to be realised in the atomic positions. 
A simple and popular method used to approximate amorphous structures is a random network model in which the atoms are placed in a lattice but distortions of both bond lengths and angles are introduced resulting in effective randomness in NNN and beyond~\cite{moss_evidence_1969,weaire_electronic_1971}. 

An approach that completely removes any underlying lattice structure is the Voronoi tessellation~\cite{florescu_designer_2009,mitchell_amorphous_2018,barghathi_phase_2014,marsal_topological_2020}.
Here, generating points are placed in a given space and polygons are defined through the areas of which any point within a given region share a closest generating point. 
The resulting graph constructed through the vertices and edges of these polygons can be used to define an amorphous configuration with $C\leq3$.
One shortcoming of this method is that the length of the edges, which correspond to interactions in the amorphous solid, are not well controlled. 
Thus, seemingly non-interacting atoms can be closer in real space than NN pairs. 
As the interaction strength is defined through atomic separation in the model we consider, Voronoi tessellation is non-physical in our setup. 
Note, a structured variation on Voronoi tessellation is given in Ref.~\cite{cassella_exact_2023}, however, while the variation of bond lengths can be reduced, the issue still persists for the interaction landscape we consider.

In this article, we propose a method to generate an amorphous solid through a gradient based variational method designed such that the separation of two atoms directly corresponds to their interaction strength. 
In particular, we propose a continuous loss functions depending on the atomic positions such that, when minimised, the output is an amorphous solid with the desired properties. 
This loss function will be composed of three terms, namely: 
(i) a first term that controls the average and variance in the coordination number of the solid; 
(ii) a second term the controls the minimum distance between atoms; 
(iii) finally, a third term that controls the maximal separation between two atoms. 
By tuning these terms we are able to readily, and efficiently, generate unique amorphous solids  of a given family of materials we are interested in defined through chosen inputs.

\subsection{Generation Protocol}

The loss function we use in this report is given by three terms, $\mathcal{L} = a_1\mathcal{L}_C + a_2\mathcal{L}_{\text{min}}+a_3\mathcal{L}_{\text{max}}$, where each $a_i$ are constant coefficients. 
In particular, 
\begin{equation}
    \begin{aligned}
        &\mathcal{L}_C = \sum_j\left|\sum_i k(r_{ij}) -m_j\right|^2,\\
        &\mathcal{L}_{\text{min}} = \sum_{ij}\left(1-\frac{1}{1+e^{-\gamma(r_{ij}-r_{\text{min}})}}\right),\\
        &\mathcal{L}_{\text{max}} = \sum_{ij}\frac{e^{r_{ij}-r_{\text{max}}}}{1+e^{-\beta(r_{ij}-r_{\text{max}})}}.
    \end{aligned}
\end{equation}
where, $r_{ij}$ is the distance between the $i^{\text{th}}$ and $j^{\text{th}}$ qubit, $k(r)$ is a Gaussian kernel in which we define the mean and variance, $m_j$ is the desired number of NNs for the $j^{\text{th}}$ atom and $\{\gamma, \beta,r_{min},r_{max}\}$ are hyper parameters. Note, the set of integers $\{m_j\}$ are picked from a normal distribution in which the mean is the desired coordination number of the solid - which itself can be a non-integer - and the variance controls the fluctuations from this coordination number.
Intuitively, the mean of the Gaussian kernel gives the desired NN separation, the variance of this kernel controls the variance in the NN separations and $r_{min}$ ($r_{max}$) is the minimum (maximum) qubit separation. 
Given this loss function, we can prepare an amorphous solid from an initial state consisting of by $N$ qubits randomly placed inside the unit square such that that no two qubits are closer than a minimum distance, $r_\text{min}$. 
We then use the PyTorch ADAM optimiser to minimise the loss~\cite{paszke_pytorch_2019,kingma_adam_2017}.

Note, the above protocol realises an amorphous solid with open boundary conditions, however, using the slightly altered approach we can also readily realise amorphous solids on a torus with periodic boundary conditions in both the $x$ and $y$ directions. 
To do so, we remove $\mathcal{L}_{\text{max}}$ from the loss function and utilise Mitchell’s best-candidate algorithm to initialise the atomic positions \cite{10.1145/127719.122736, cassella_exact_2023}. 
See Appendix~\ref{app:examples} for examples of amorphous solids generated through this variational protocol with both types of boundary conditions.
Throughout this work we will work with systems with open boundaries as they are more readily studied experimentally.

\subsection{Structure Factor}

Typically, the precise positions of atoms in an amorphous solid are not known. 
However, the static structure factor - the Fourier transformation of the atomic positions - can be measured in scattering experiments. 
This results can then be used to diagnose properties of the solid in consideration. 
In particular, the radius of this circle reflects well defined average NN separation, and by computing the Hankel transform we can approximate the radial distribution function and by extension the coordination number \cite{PhysRevE.94.030701}.

As a verification that the solids generated we calculate their static structure factor. 
In regular lattices the structure factor contains sharp Bragg peaks at regular intervals due to the long-range order of these materials.
However, in an amorphous layout such peaks do not appear. 
In fact, we observe a circle surrounding the origin due the isotropic nature of amorphous materials~\cite{zallen_physics_1998}. 

\begin{figure}
    \centering
    \includegraphics[width=\columnwidth]{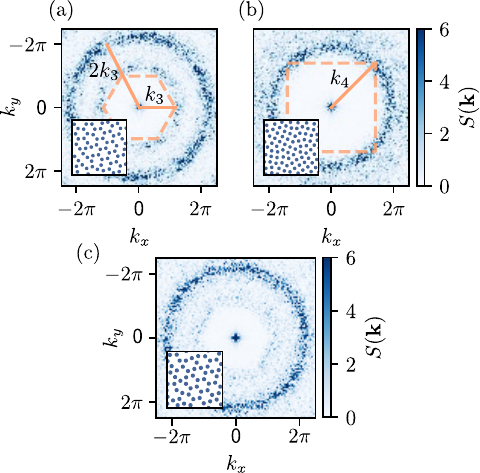}
    \caption{The static structure factor of amorphous solids with coordination number $C\approx3$ (a) and $C\approx4$ (b) and $C\approx3.5$ (c). The inset of all panels shows a patch of each amorphous solid. In panels (a) and (b) we highlight the correspondence to the Brillouin zone of the regular lattice with the equivalent coordination number for each case - honeycomb for $C=3$ and square for $C=4$ - by plotting this in orange. Furthermore, we show the reciprocal lattice vectors, $k_3$ and $k_4$, for these regular lattices. Clearly these vectors predict the radius of both circles that appear in each amorphous solid structure factor.}
    \label{fig:amorphous}
\end{figure}

In Fig.~\ref{fig:amorphous}(a), (b) and (c) we present the static structure for an amorphous solid with $C\approx3$, $C\approx4$ and $C\approx3.5$ respectively; 
representative sections of the real-space atomic positions are included in this figure as insets. 
In each of these plots we clearly see the expected circular pattern in the structure factor. 
Furthermore, for the cases $C\approx3$ and $C\approx4$ we outline the Brillouin zone for the regular lattice that shares the the coordination number of each amorphous material, i.e., the honeycomb lattice for $C=3$ and the square lattice for $C=4$.
By ensuring that the each regular lattice has the same NN separation as the average NN separation of the respective amorphous solid, we are able to predict the radius of the circles in the static structure factor in both cases. 
Note, the structure factor of the $C\approx3.5$ solid is expected to be similar to that of the $C\approx3$ solid as it mainly consists of hexagonal and kagome-type palettes and thus share the same reciprocal lattice vectors.

\section{\label{sec:lswt}Review of Linear Spin Wave Theory}
Solving the Hamiltonian~\eqref{eq:ising_hamiltonian} for the spin positions $\mathbf{r}_i$ of a quantum amorphous solid represents a hard numerical task. In this regard, note that, despite the well-known exponential increase of computational needs with the system size $N$, many equilibrium properties of 2D regular spin lattices have been studied with state-of-the-art techniques such as sparse state vector methods~\cite{fehske_exact_2008}, quantum Monte Carlo, or tensor network approaches~\cite{verstraete_matrix_2008,orus_practical_2014} targeting the relevant many-body Hilbert space region for the ground state. While using these methods for non-equilibrium studies generally represents an even harder numerical task, they can be still used to benchmark the unitary dynamics of finite systems at shorter times. 
Both in the equilibrium and non-equilibrium case, the translational symmetry of regular lattices typically allows one to extract relevant physics from finite-size scaling of relatively small systems, or to reduce the number of degrees of freedom by assuming a repeated unit cell~\cite{schollwock_density-matrix_2011,mcculloch_infinite_2008,jahromi_universal_2019} commensurate with the lattice. Moreover, in the context of tensor networks, the lattice provides a natural starting point to find the optimal topology of the network minimising the amount of entanglement needed to simulate the properties of the material~\cite{naghmouchi_optimization_2024}. In stark contrast, amorphous solids can only be properly defined for large $N$ due to their characteristic lack of long-range order, which in turn prevents a natural choice for a network topology or a unit cell. In view of these numerical challenges, here we use a perturbative approach, the linear spin wave theory (LSWT),  
which can be used to benchmark the nature of the excitations in disordered 2D spin systems~\cite{vojta_excitation_2013,alvarez_zuniga_bose-glass_2013,menu_anomalous_2020}. In particular, within the LSWT approximation we can
characterize part of the phase diagram and linear excitations of amorphous solids at very large system sizes $N\approx 1000$. While this analysis provides a first intuition of the quantum regime  in amorphous spin systems, general phase diagrams and dynamical properties of quantum amorphous solids should be studied with more advanced numerical methods, or quantum simulators.

The LSWT (see  Appendix~\ref{app:lswt} for a an extended discussion) relies on the approximation of highly polarized semi-classical spins satisfying $\langle\tilde{S}^z_i\rangle\approx 1/2$ in a proper rotated basis given by
\begin{equation}
\begin{split}
S^z_i&=\tilde{S}^z_i\cos\theta_i+\tilde{S}^x_i\sin\theta_i, \\
S^x_i&=\tilde{S}^x_i\cos\theta_i-\tilde{S}^z_i\sin\theta_i,
\end{split}
\end{equation}
where the optimal angles $\theta_i$ are found via minimization of the semi-classical mean-field energy of the system. Within this mean-field approximation, one can express 
the mean-field ferromagnetic order parameter as
\begin{equation}\label{eq:M}
    M= \frac{1}{N}\sum_i \langle S^z_i\rangle \simeq\frac{1}{2N}\sum_i \cos\theta_i.
\end{equation}
The LSWT considers excitations on top of this mean-field phase. That is, under the assumption of harmonic quantum fluctuations, the spins can be bosonized around their classical value following the Holstein-Primakoff approximate mapping~\cite{holstein_field_1940} $\tilde{S}^z_j = \frac{1}{2}-a\dagg_ia_i,\ \tilde{S}^x_j = \frac{1}{2}(a\dagg_i+a_i)$, where $a_i\ (a\dagg_i)$ are anihilation (creation) bosonic operator satisfying $[a_i,a\dagg_j]=\delta_{ij}$. After the abovementioned approximations, $H$ can be written as the quadratic bosonic form
\begin{equation}
\begin{split}
    H =&-\frac{1}{4}\sum_{i<j} J_{ij} \sin\theta_i\sin\theta_j (a\dagg_i+a_i)(a\dagg_j+a_j)\\
    &+\sum_i \left[h_x\sin\theta_i+\frac{\cos\theta_i}{2}\sum_j J_{ij}\cos\theta_j\right]a\dagg_ia_i.
    \end{split}
\end{equation}
The last step in the LSWT approach consists on diagonalizing the Hamiltonian by definig the so called Bogoliubov modes $a_i=\sum_\mu u_{i,\mu}b_\mu+v_{i,\mu}b\dagg_\mu$. In terms of the bosonic Bogoliubov modes the Hamiltonian takes the diagonal form 
\begin{equation}
    H =\sum_{\mu=1}^N \omega_\mu b\dagg_\mu b_\mu + E_g,
\end{equation}
where the frequencies $\omega_\mu$ constitute the LSWT spectrum of the system, and $E_g$ is a global energy shift. One can then access relevant quantities of the original problem. 
For instance, the bulk gap of the LSWT spectrum is given by
\begin{equation}
    \Delta=\min\{ \omega^\text{bulk}_\mu\},
\end{equation}
where $\{\omega^\text{bulk}_\mu\}$ is the set of energies corresponding to Bogoliubov modes with a finite support in the bulk of the system (see details in the Appendix~\ref{app:lswt}).
\begin{figure}[t]
    \centering
\includegraphics[width=\columnwidth]{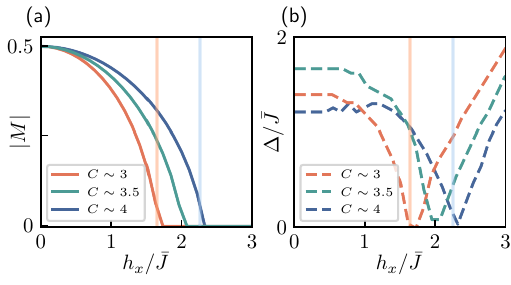}
    \caption{Evolution of (a) the mean-field ferromagnetic order parameter $|M|$ and (b) the LSWT bulk gap  $\Delta$ as a function of the transverse field $h_x$ for amorphous solids with different coordination numbers. Here we fix $N=400$ and a power-law interaction decay of $\alpha=6$. The vertical blue (orange) line correspond to the transition point predicted by LSWT in the square (honeycomb) regular lattice, which has a coordination number of $C=4$ ($C=3$).}
    \label{fig:pd_lswt}
\end{figure}
\section{\label{sec_mf_pd}Mean-field phase diagram for \texorpdfstring{$J_0>0$}{J0>0}}
We start by discussing the phase diagram of $H$ as a function of both the ratio $h_x/\bar{J}$ and the properties of the amorphous solid encoded in $\mathbf{r}_i$. For simplicity, we consider ferromagnetic interactions ($J_0>0$) to avoid frustrated antiferromagnetic phases that cannot be properly studied with LSWT due to the presence of a large amount of entanglement. Figure~\ref{fig:pd_lswt} shows the relevant LSWT quantities as a function of $h_x/\bar{J}$ across the ferromagnet to quantum paramagnet phase transition for amorphous solids with $C\approx 3$,  $C\approx 3.5$, and $C\approx 4$. In particular, we show the mean-field ferromagnetic order parameter $|M|$, and the LSWT energy gap $\Delta$, which is expected to close for a continuous quantum phase transition. By comparing the behavior of the three coordination numbers, we note that, as expected, increasing this parameter leads to a larger interaction energy scale, and shifts the critical point to a larger $h_x/\bar{J}$. While the phase diagram is qualitatively similar to that of regular lattices (see vertical lines in Fig.~\ref{fig:pd_lswt}) we note two main differences for the amorphous solid case. First, even if we fix the same NN coordination number and average interaction strength $\bar{J}$, the critical point will in general differ from that of a regular lattice, due to the difference between longer-range interactions under a fixed power-law decay. Second, while regular lattices are restricted to integer coordination numbers, the coordination number of an amorphous solid can also take fractional values, leading to a $C\approx 3.5$ critical point that interpolates between the transition points of two integer coordination numbers. As a final comment, note that the mean field approximation used to compute the transition points in Fig.~\ref{fig:pd_lswt} in general overestimates the ferromagnetic order. For instance, for the paradigmatic 2D transverse field Ising model in the square lattice with only NN interactions, the mean-field transition point is at $h_x=2J$, while it is known from more accurate numerical methods that the true transition occurs at $h_x\approx 1.5J$~\cite{blote_cluster_2002,orus_simulation_2009}. 

\section{Delocalized nature of the spectrum for \texorpdfstring{$J_0>0$}{J0>0}}\label{sec_loc}
\begin{figure}[b]
    \centering
    \includegraphics[width=\columnwidth]{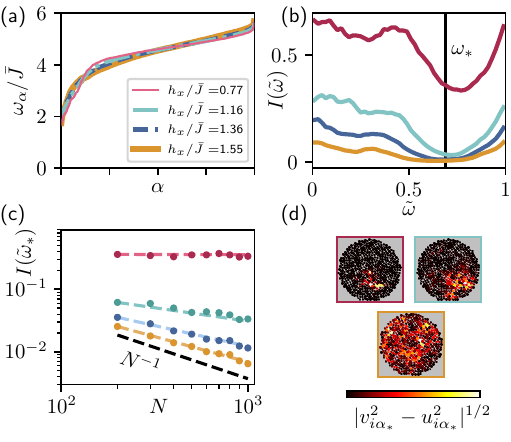}
    \caption{Properties of the LSWT spectrum for an amorphous solid with $C\approx 4$ and for three different values of $h_x/\bar{J}$, the legend indicates the color code used in all the panels. (a)-(b) Energy dispersion and inverse participation ratio for $N=1000$. (c) Scaling of $I$ with the system size at a fixed energy $\omega_*(h_x)$, indicated as a vertical line in (b) for the largest $h_x$. (d) Spatial profile of a mode with $\omega_\alpha=\omega_*$, for three values of $h_x$ and $N=500$.}
    \label{fig:localization_lswt}
\end{figure}

Let us now investigate the nature of the LSWT excitations in the ferromagnetic phase of amorphous solids described by the Ising model of Eq.~\eqref{eq:ising_hamiltonian}. In the classical ferromagnet limit $h_x/J_0\to 0$, the Bogoliubov modes coincide with the local modes, i.e., $b_i=a_i$. If all these local modes have the same local energy, as e.g., in a regular square lattice without disorder, a finite transverse field $h_x$ leads to a strong mode hybridization and a delocalized dispersive band. Local disorder competes with such a delocalization mechanism, as it induces energy shifts in the local mode energies, effectively preventing hybridization. As an example of such a disorder-induced localization in a similar model, the authors of Ref.~\cite{menu_anomalous_2020} recently considered the effect of local disorder in the  positions $\mathbf{r}_i$ of a regular square lattice. Due to the strong dependence of the interaction with position through the $|\mathbf{r}_i-\mathbf{r}_j|^6$ decay, it was shown that, for a fixed $h_x=\bar{J}$, the LSWT spectrum exhibits a smooth transition from delocalized modes at positional disorders below $3 \%$, to fully localized modes when disorder is $5\%$ or larger.
It is thus natural to investigate what are the localization properties of the LSWT spectrum in amorphous solids, which bear some similarities both with regular lattices and disordered systems. For this analysis, we work with amorphous solids with $C\approx 4$, and sizes up to $N=1000$. The results are presented in Fig.~\ref{fig:localization_lswt}, where we compare the behavior at different transverse fields $h_x$.
Note that, for these $h_x$, the system is still deep in the mean-field ferromagnetic phase, with $|M|\approx 0.7 |M|_\text{max}$ at most.  

Figure~\ref{fig:localization_lswt}(a) shows that the dispersion of the LSWT spectrum is apparently similar for the different values of $h_x$. We observe that larger $h_x$ lead to slightly higher dispersion at the center of the band. The strong dispersion close to the band minima is related to the open boundary condition, which leads to in-gap states that have lower energy than the bulk ones. The localization properties of these LSWT spectra can be studied by means of the inverse participation ratio (IPR) defined as
\begin{equation}
    \text{I}(\tilde{\omega})\equiv \frac{1}{\mathcal{N}(\tilde{\omega},\epsilon)}\sum_{\abs{\tilde{\omega}_\alpha-\tilde{\omega}}<\epsilon,i}\left[(v_{i\alpha})^2-(u_{i\alpha})^2\right]^2,
\end{equation}
where $\tilde{\omega}_\alpha=(\omega_\alpha-\omega_\text{min})/(\omega_\text{max}-\omega_\text{min})\in [0,1]$ is the normalized  LSWT spectrum, $\epsilon$ provides a discrete energy window (here we set $\epsilon=0.02$), and $\mathcal{N}(\tilde{\omega},\epsilon)$ counts the number of modes within this energy window around $\tilde{\omega}$. $\text{I}(\tilde{\omega})$ thus gives the average inverse volume of the Bogoliubov modes around the normalized energy $\tilde{\omega}$. In particular, $I(\tilde{\omega})$ tends to a finite value in the thermodynamic limit for localized modes, while it goes to zero as $N^{-1}$ for fully extended modes. 

In Fig.~\ref{fig:localization_lswt}(b), we can observe that for the smaller field $h_x$ the modes are localized, as $I(\tilde{\omega})$ is clearly different from zero and the system is already close to the thermodynamic limit. For increasing $h_x$, we observe a smooth transition to delocalized modes around the most delocalized frequency $\omega_*$. The edge frequencies remain localized, as they are related to edge sites or bulk defects with a different coordination number than the average.  To rigorously certify the localization properties of the modes, in Fig.~\ref{fig:localization_lswt}(c) we show the finite-size scaling of $I(\tilde{\omega_*})$, see vertical lines in Fig.~\ref{fig:localization_lswt}(b).  We observe that $I(\tilde{\omega_*})$ remains finite for small $h_x/\bar{J}$, as it does not decay with the system size, consistent with a fully localized spectrum. For $h_x/\bar{J}\gtrsim 1 $, we observe quasi-delocalized modes with $I(\tilde{\omega_*})$ decaying slower than the fully delocalized scaling $N^{-1}$. Further increasing $h_x$ leads to a smooth transition to an approximate $N^{-1}$ scaling, consistent with almost fully delocalized modes. Finally, in Fig.~\ref{fig:localization_lswt}(d) we show the spatial profile of mode instances at $\omega_*$ for different $h_x$, from which one can clearly visualize the smooth evolution from localized to delocalized modes in a real space picture.

\section{Dynamical structure factor for \texorpdfstring{$J_0>0$}{J0>0}}\label{sec_dsf}
The excitation properties inside the ferromagnetic phase of an amorphous solid can also be studied in momentum space by computing dynamical structure factors. For the LSWT eigenmodes, these take the simple form
\begin{equation}\label{eq:dsf_zz}
    S^{zz}(\tilde{\omega}, \mathbf{k})=\sum_{\abs{\tilde{\omega}_\alpha-\tilde{\omega}}<\epsilon} \left[\sum_i e^{-i \mathbf{k}\cdot\mathbf{r}_i}(u_{i,\alpha}+v_{i,\alpha})\frac{\sin\theta_i}{\sqrt{4N}}\right]^2,
\end{equation}
\begin{equation}\label{eq:dsf_xx}
    S^{xx}(\tilde{\omega}, \mathbf{k})=\sum_{\abs{\tilde{\omega}_\alpha-\tilde{\omega}}<\epsilon} \left[\sum_i e^{-i \mathbf{k}\cdot\mathbf{r}_i}(u_{i,\alpha}+v_{i,\alpha})\frac{\cos\theta_i}{\sqrt{4N}}\right]^2.
\end{equation}
For regular lattices with large disorder, which strongly breaks translational invariance, the momentum $\mathbf{k}$ ceases to be a good quantum number. As a consequence, the dynamical structure factors are expected to exhibit finite momentum widths related to the appearance of localized modes. In the case of amorphous solids, while translational invariance is clearly broken, the fact that there is a well-defined coordination number can lead to much richer properties. In this regard, note that despite the strong breaking of translational invariance in amorphous solids, we have already seen in the previous sections that their static structure factor exhibits well-defined rings in momentum space, and that the LSWT spectrum is  delocalized for sufficiently large $h_x$. 
\begin{figure}[t]
    \centering
    \includegraphics[width=\columnwidth]{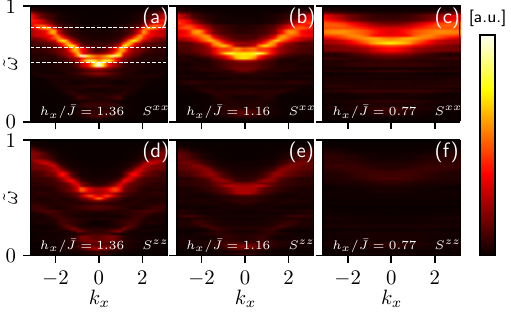}
    \caption{Dynamical structure factors for an amorphous solid with $N=1000$ and $C\approx 4$ at a fixed $k_y=0$. The upper panels (a)-(c) show the transverse dynamical structure factor $S^{xx}$, while the lower panels (d)-(f) correspond to the longitudinal one $S^{zz}$. The value of transverse field $h_x/\bar{J}$ is indicated on each plot.}
    \label{fig:dsf_band}
\end{figure}

Figure~\ref{fig:dsf_band} shows the dynamical structure factors of an amorphous solid with $N=1000$ and $C\approx 4$. In particular, here we consider the dispersion relation given by $S^{\mu \mu}(\tilde{\omega}, k_x)$ at a fixed $k_y=0$ and for three values of the transverse field $h_x/\bar{J}$, already studied in Fig.~\ref{fig:localization_lswt}.  A first comment is that, for the ferromagnetic phase under consideration, which is polarized close to the $z$ axis, the transverse component $S^{xx}$, shown in Figs. ~\ref{fig:localization_lswt}(a)-~\ref{fig:localization_lswt}(b), is generally larger than the longitudinal one $S^{zz}$, shown in Figs.~\ref{fig:localization_lswt}(d)-~\ref{fig:localization_lswt}(f). This can be readily inferred from the trigonometric factors in Eqs.~\eqref{eq:dsf_zz}-\eqref{eq:dsf_xx}. Besides this, in all the panels we can 
identify a bright bulk band whose minima is located in the middle of the normalized LSWT spectrum, and a fainter band at low energy, mainly due to open boundary excitations. The effect of the transverse field is consistent with our analysis of Fig.~\ref{fig:localization_lswt}. In Figs.~\ref{fig:dsf_band}(a) and \ref{fig:dsf_band}(d), the large transverse field $h_x/\bar{J}=1.36$ leads to a localized bulk band in momentum space, which implies delocalized real-space excitations. For an intermediate value $h_x/\bar{J}=1.16$, see Figs.~\ref{fig:dsf_band}(b) and \ref{fig:dsf_band}(e), the bulk band acquires a larger momentum width, implying less delocalized real-space modes. Finally, for the smallest transverse field $h_x/\bar{J}=0.77$ the band is clearly delocalized within the bandwidth in momentum space, signalling highly localized real-space excitations.

\begin{figure}[t]
    \centering
    \includegraphics[width=\columnwidth]{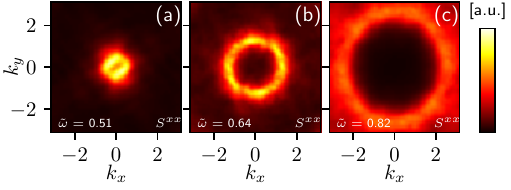}
    \caption{Dynamical structure factor $S^{xx}$ for $h_x/\bar{J}=1.36$. Each panel corresponds to a different frequency cut, indicated in each plot.}
    \label{fig:dsf_cuts}
\end{figure}

In order to further characterize the bulk band unveiled by the dynamical structure factor, in Fig.~\ref{fig:dsf_cuts} we show the transverse component $S^{xx}(\tilde{\omega},\mathbf{k})$ in the delocalized regime at three different cuts of increasing frequency, corresponding to the dashed lines of Fig.~\ref{fig:dsf_band}(a). In the three panels, we can clearly observe that the dynamical structure factor exhibits rotational invariance in momentum space, i.e., the band energy cuts take the form of rings. Note that, while these resemble the ring  observed in the static structure factor (Fig.~\ref{fig:amorphous}), here we are dealing with dynamical properties, and thus the dispersion translates into a ring radii that increases with energy. A final comment is in order concerning the effect of the frequency on the localization of the rings. The momentum width is relatively small at the band minima and it becomes particularly narrow in the middle of the band, corresponding to the most delocalized spectral  region. In contrast, the higher energy ring of Fig.~\ref{fig:dsf_cuts}(c) is delocalized in momentum space, as the band hybdridizes with localized states at the edge of the bulk spectrum.  
Interestingly, this behavior complements the analysis of Fig.~\ref{fig:localization_lswt}(b), where one can also identify the most delocalized spectral region around $\tilde{\omega}^*$.

\section{Antiferromagnetic case \texorpdfstring{$J_0<0$}{J0<0}: geometrical frustration and disorder}\label{sec:simulated_annealing}

For antiferromagnetic interactions $J_0<0$, spin glass physics~\cite{edwards_theory_1975, binder_spin_1986} can naturally arise in the model of Eq.~\eqref{eq:ising_hamiltonian}, since an amorphous magnet exhibits in general both frustrated and unfrustrated plaquettes. Note that this is in stark contrast with regular lattices, in which at $T=0$ and zero transverse field $h_x=0$, one has either an unfrustrated geometry leading to a long-range antiferromagnetic state (e.g., in the square lattice), or a frustrated geometry leading to a spin liquid (e.g., in the triangular or kagome lattices). Interestingly, such a scenario of spatially inhomogeneous magnetic frustration bears strong similarities with paradigmatic models of 2D spin glasses, such as the Edwards-Anderson (EA) model~\cite{edwards_theory_1975} in the square lattice, in which a randomized NN couplings with a bimodal ($\pm J$) or Gaussian distribution leads to a 2D spin glass at $T=0$~\cite{kawashima_critical_1992, campbell_energy_2004,houdayer_low-temperature_2004,rieger_critical_1996,rubin_dual_2017,xu_dynamic_2017}. The main difference, however, is that, in an amorphous solid, glassiness arises from the interplay between antiferromagnetic interactions and the real-space geometry~\cite{kawamura_simplified_1983}, rather than the competition between antiferromagnetic and ferromagnetic couplings. For instance, for $C=4$ we can generate amorphous layouts with large subregions that are similar to an square or kagome lattice, with completely different properties at $T=0$. Moreover, we can tune the coordination number to intermediate values without a regular lattice analog, such as e.g., $C=3.5$ or $C=5$. 

While, therefore, antiferromagnetic amorphous magnets are expected to exhibit a rich low-energy landscape, the numerical benchmark of these systems in the quantum regime is beyond the scope of this work, since it is much more complex than in the ferromagnetic case. In particular, the absence of a long-range ordered classical reference state prevents the perturbative LSWT analysis of the previous sections. Instead, in this section we focus on the analysis of the classical low-energy landscape of amorphous antiferromagnets, which by itself represents a complex numerical task. To this aim, we use simulate annealing (SA)~\cite{kirkpatrick_optimization_1983}, a simple method that illustrates the complexity of antiferromagnetic amorphous magnets, and some common features with both regular lattices and spin glass models. Nevertheless, we note that it would also be interesting to perform this classical analysis with more sophisticated methods, such as parallel tempering Monte Carlo sampling~\cite{swendsen_replica_1986,wang_comparing_2015,yucesoy_correlations_2013,janus_collaboration_critical_2013} in the limit $T\to 0$, or optimization methods for finding classical ground state energies and configurations of spin systems~\cite{thomas_matching_2007,hartmann_ground_2011}. 

\subsection{Review of simulated annealing}
In order to provide a first benchmark of the complex antiferromagnetic energy landscape of the amorphous layouts studied in this work, we consider the classical low-energy states obtained through SA of classical Ising magnets described by the limit $h_x\to 0 $ of the Hamiltonian~\eqref{eq:ising_hamiltonian}
\begin{equation}\label{eq:classical_ising_hamiltonian}
H_\text{classical}=-\sum_{i<j}J_{ij}\sigma_i\sigma_j,\ \ \sigma_i=\pm 1.
\end{equation}
In particular, for a given layout $J_{ij}$, we anneal $N_R=60$ statistically independent replicas, hereafter labeled by the index $\alpha$, during $n_\text{steps}$ Monte Carlo sweeps. Each Monte Carlo sweep consists in flipping $N$ randomly selected spins, according to the Metropolis update rule. After each Monte Carlo sweep, we decrease the temperature according to 
\begin{equation}
    T_i = T_0\left(1-\frac{i}{n_{\text{steps}}}\right),\ i=1,\dots,n_\text{steps}.
\end{equation}
A first quantity of interest is then the final classical energy of the different annealed replicas
\begin{equation}
    E^\alpha_\text{SA} \equiv \langle H_\text{classical}\rangle_\alpha ,
\end{equation}
which corresponds to the ground-state energy in the limit of an infinitely slow annealing time, i.e., $n_\text{steps}\to \infty$.

We are also interested in the EA parameter, defined as the averaged spin overlap between two replicas $\alpha$ and $\beta$ after SA
\begin{equation}
\begin{split}
    q_{\text{SA}}^2 &\equiv \frac{1}{N_R(N_R-1)}\sum_{\alpha \neq \beta} |q_\text{SA}^{\alpha \beta}|^2,\\
    q_\text{SA}^{\alpha \beta}&\equiv \frac{1}{N}\sum_i^N \sigma^{(\alpha)}_i\sigma^{(\beta)}_i,
    \end{split}
\end{equation}
and its density distribution $P(q_\text{SA}^{\alpha \beta})$. To compute this quantity, we reduce the set of 60 replicas, and only use the 20 replicas with lower final energy. Note that, the form of $P(q_\text{SA}^{\alpha \beta})$ provides a lot of information about the low-energy states of the system, and it is directly related to the replica symmetry breaking picture of the spin glass phenomenon.

\subsection{Simulated annealing in amorphous solids}
\begin{figure}[h]
  \includegraphics{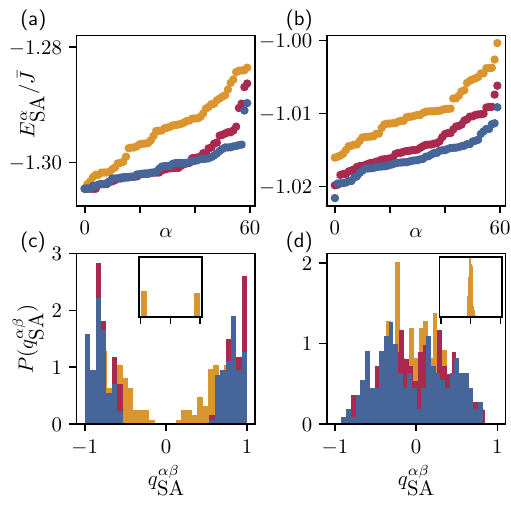}
  \caption{SA results for a square-like amorphous solid, panels (a) and (c), and a kagome-like amorphous solid, panels (b) and (d), both with $N=400$. Yellow, purple, and blue colors are used for $n_\text{steps}=10^5,\,10^6,\,$ and $10^7$, respectively. Panels (a) and (b) show the final SA energies of the $N_R=60$ statistically independent replicas. Panels (c) and (d) show the probability distribution of the overlap between the 20 lowest-energy replicas. The insets in (c) and (d) correspond to the regular square ($N=400$) and kagome ($N=432$) lattices replica overlaps in the ground state, see Appendix~\ref{app:SA_results} for details.}
  \label{fig:simulated_annealing_amorphous}
\end{figure}

Figure~\ref{fig:simulated_annealing_amorphous} shows the SA results corresponding to two amorphous solids with $N=400$, shown in Appendix~\ref{app:SA_results}. In particular, here we consider an amorphous solid with $C\approx 4$ with local NN plaquettes resembling the ones of a square lattice, and another amorphous solid with $C\approx 4$ and local NN plaquettes resembling the ones of a kagome lattice. 

A first general remark is that, for the large system size under consideration, the glassy nature of the low-energy landscape manifests itself in the fact that the SA is not able to find the ground state of the system. That is, as shown in Figs.~\ref{fig:simulated_annealing_amorphous}(a)-(b), the final energies $E^\alpha_\text{SA}$ still depend on the annealing time at the largest annealing times considered here $n_\text{steps}=10^6-10^7$. As shown in Appendix~\ref{app:SA_results}, a similar lack of convergence is observed for EA models at comparable system sizes. The latter is in agreement with the picture that the antiferromagnetic amorphous layouts lead to energy landscapes of comparable complexity to those of models with ferromagnetic and antiferromagnetic couplings.   

The lack of convergence of the SA also implies that the replica overlaps and their distributions, depicted in Figs.~\ref{fig:simulated_annealing_amorphous}(c)-(d), do not longer correspond to ground states, but rather to the overlaps of low-energy annealed solutions. Still, the distributions $P(q_\text{SA}^{\alpha \beta})$ shown in Figs.~\ref{fig:simulated_annealing_amorphous}(c)-(d) suggest that there are strong differences between the low-energy landscapes of the square and kagome-like amorphous solids, despite both having the same $C\approx 4$ and lacking long-range (lattice) order. The square-like amorphous solid exhibits two peaks at large $|q_\text{SA}^{\alpha \beta}|$, reminiscent of the square lattice antiferromagnet delta functions, shown in the inset. The value of $q_\text{SA}^2\approx 0.45$, for the largest annealing time, also suggests a strong spin-glass order, despite clearly deviating from a unit value, due to the frustration of a long-range antiferromagnetic unit cell. In contrast, the kagome-like amorphous solid exhibits a finite but small spin glass order parameter $q_\text{SA}^2\approx 0.11$ for the largest annealing time, as $P(q_\text{SA}^{\alpha \beta})$ has a single wide peak structure centered at $q_\text{SA}^{\alpha \beta}=0$. Again, the latter is reminiscent of the behavior observed in the regular kagome lattice, as shown in the inset.

In conclusion, these SA results support, on the one hand, that for antiferromagnetic couplings, amorphous solids exhibit a complex classical low-energy landscape, similar to paradigmatic spin-glass models at comparable system size (see Appendix~\ref{app:SA_results}). On the other hand, they also suggest that the interplay between the coordination number and the local bond angle distribution can have a large large impact on the properties of this landscape, with a behavior that can be partially explained by that of parent regular lattices.  
In this context, a quantum annealing protocol could potentially help in unveiling the low-energy classical landscape of amorphous magnets. Furthermore, it would also be very interesting to take into account the effect of quantum fluctuations in the finite  $h_x/J_0$ phase diagram and study, for example, the competition between the glassy phase and the potential spin liquid phase. This is however extremely challenging for TN and QMC numerical methods, due to the large system sizes and the presence of frustration already at the classical level. In the next section, we discuss how such a complex antiferromagnetic case can be studied with a quantum simulator.
 
 \section{\label{sec:experiment}Quantum simulation of amorphous solids}
The Hamiltonian~\eqref{eq:ising_hamiltonian} can be naturally simulated in arrays of optical tweezers loaded with Rydberg atoms~\cite{henriet_quantum_2020,browaeys_many-body_2020}, which exhibit large dipole-dipole interactions~\cite{saffman_quantum_2010}. In particular, the use of programmable moving optical tweezers during the loading stage is highly suited to deterministically trap individual atoms at positions $\mathbf{r}_i$ corresponding to different amorphous solids in the 2D plane. Moreover, large systems with $N> 300$ have been already realized in this platform~\cite{schymik_situ_2022}, and larger systems are expected to become available in the near future. 
 \subsection{Hamiltonian engineering}
 By considering the local qubit
 consisting on the atomic ground state $\ket{0}$ and a given Rydberg state $\ket{1}$, the Hamiltonian of the simulator taking into account the effect of the Rydberg interaction and an external laser can be expressed as 
\begin{equation}\label{eq:rydberg_hamiltonian}
H_{\text{Ryd}}=\sum_{i<j}\frac{C_6}{|\mathbf{r}_i-\mathbf{r}_j|^6}n_in_j+\frac{\hbar\Omega(t)}{2}\sum_i \sigma^x_i-\sum_i\frac{\hbar\delta_i(t)}{2} \sigma^z_i,
\end{equation}
where $n_i=(1+\sigma^z_i)/2$, $C_6>0$ is the strength of the Van der Waals interaction due to the dipole-dipole Rydberg interaction, and $\Omega(t)$ and $\delta_i(t)$ are the Rabi frequency and local detuning induced by the external laser. By setting $C_6=|J_0|$, and assuming time-independent laser parameters $\hbar\Omega=h_x$ and $\hbar\delta_i=\sum_{j}J_{ij}/2$, one can readily see that the Hamiltonians of Eqs.~\eqref{eq:ising_hamiltonian} and ~\eqref{eq:rydberg_hamiltonian} become equivalent up to an irrelevant constant. It is worth noticing that, despite $H_\text{Ryd}$ always consists of antiferromagnetic interactions ($J_0<0$), we can also probe the dynamics of the ferromagnetic regime, as discussed below.

\subsection{Adiabatic state preparation and unitary evolution}
In the experimental protocol, one typically assumes that the initial state is the atomic groundstate $\ket{0}^{\otimes N}$, which corresponds to a fully polarized ferromagnet in the spin language. For a very large initial detuning $\delta>0$ or $\delta<0$, the atomic groundstate is the non-degenerate groundstate or highest excited state of $H_\text{Ryd}$, respectively. Following the adiabatic theorem, one can then use an initial large negative (positive) detuning to prepare the groundstate (highest excited state) of $H_\text{Ryd}$ by slowly bringing the detuning and Rabi frequency to the final desired value. Interestingly, this means that, despite the fixed sign of the interactions $J_0<0$ in the abovementioned setup, one can adiabatically prepare both antiferromagnetic and ferromagnetic phases.
After preparing an eigenstate of $H_\text{Ryd}$, the quantum simulator can also be used to study the unitary evolution of the system after a sudden quantum quench in which the Hamiltonian is abruptly changed from $H_\text{Ryd}\to H_\text{Ryd}'$. 
Concerning such unitary dynamics, 
 note that for initial states preserving time-reversal symmetry, and under the condition $\hbar\delta_i=\sum_{j}J_{ij}/2$, the evolutions under $H_{\text{Ryd}}$ and $-H_{\text{Ryd}}$ are indistinguishable, which means that the sign of $J_0$ is irrelevant for states with real coefficients in the computational basis. 

 \subsection{Detection schemes}
As the final step of the quantum simulation, one can access relevant microscopic quantities of the system. In particular, the local magnetization $\langle\sigma^z_i\rangle$ and its moments (e.g.,$\langle\sigma^z_i\sigma^z_j\rangle$) can be directly accessed via fluorescence images of the final state, which for each experimental run provides a snapshot with single-site resolution of the atomic populations $n_i$. From this, one can compute local order parameters of magnetic phases (e.g., Eq.~\eqref{eq:M} for the ferromagnetic case) and study quantum phase transitions in amorphous solids. 
Concerning the properties of excitations, they can be typically investigated through their impact in the spectral properties of the post-quench unitary dynamics. In this direction, recent works discussed the possibility of measuring dynamical structure factors with quantum simulators~\cite{knap_probing_2013,baez_dynamical_2020,li_detecting_2022}.

\section{\label{sec:conclusion}Conclusion and Outlook}

In this work, we investigated the properties of quantum amorphous magnets that could be simulated in state-of-the-art Rydberg atom arrays. 
To this end, we first developed a variational protocol that generates well-defined amorphous configurations for systems with quasi long-range Ising interactions, i.e., decaying as $1/r^6$. 
We then used perturbative techniques to benchmark the physics of these newly generated Hamiltonians assuming finite but small quantum fluctuations. We started by studying their mean-field ferromagnetic phase transitions, for which we revealed that the effect of the coordination number in amorphous magnets is similar as in regular lattices. We further unveiled the nature of linear spin wave excitations inside the ferromagnetic phase. By computing the inverse participation ratio, we observed that, in contrast to fully disordered systems, this spectrum can still be delocalized in amorphous magnets with a moderate transverse field. This result was supported by the behavior of the dynamical structure factors, which exhibited full rotational symmetry, and that were consistent with a dispersive band in the delocalized regime. Importantly, note that our perturbative analysis fails to describe regimes with large quantum fluctuations, which prevented us to study quantum critical regions accurately, or to consider the antiferromagnetic quantum regime, where frustrated amorphous magnets are expected to arise. In this antiferromagnetic case, we have carried out a simulated annealing analysis of the model in its classical Ising limit, which indeed unveiled a complex low-energy classical landscape. A natural research line would thus be to study the interplay between such classical magnetic frustration and quantum fluctuations.  With the motivation to tackle these more challenging regimes with the help of a quantum simulator, we proposed an experimental setup based on neutral atoms trapped in programmable optical tweezers, and laser-coupled to their Rydberg states, to realize the same class of amorphous magnet Hamiltonians studied in this work. In particular, we described how this Hamiltonian could be engineered in a platform with $N>100$ qubits, the possibility to adiabatically prepare ground states or perform post-quench unitary dynamics, and the available detection schemes. This opens the road to study amorphous solids described as a collection of interacting two-level systems in regimes that are difficult to simulate classically. In this regard, note that some low-temperature properties of amorphous solids have been already explained by models based on noninteracting two-level systems~\cite{phillips_tunneling_1972,anderson_anomalous_1972}, but there are still many open questions concerning the nature of the two-level systems, and the effect of interactions~\cite{coppersmith_frustrated_1991,faoro_interacting_2015, arnold_experimental_1975, burin_interactions_1998, carruzzo_nonequilibrium_1994,agarwal_polaronic_2013,muller_towards_2019, lisenfeld_observation_2015}.

\acknowledgments
We thank R. Menu and T. Roscilde for fruitful discussions on the linear spin wave theory. We also thank V. Elfving, J. Knolle, and T. Mendes-Santos for insightful discussions. AD and JV acknowledge funding from the European Union under Grant Agreement 101080142 and the project EQUALITY. The authors would like to thank the Institut Henri Poincaré (UAR 839 CNRS-Sorbonne Université) and the LabEx CARMIN (ANR-10-LABX-59-01) for their support.

\appendix
\section{Examples of amorphous solids}\label{app:examples}
\begin{figure}
    \centering
    \includegraphics[width=\columnwidth]{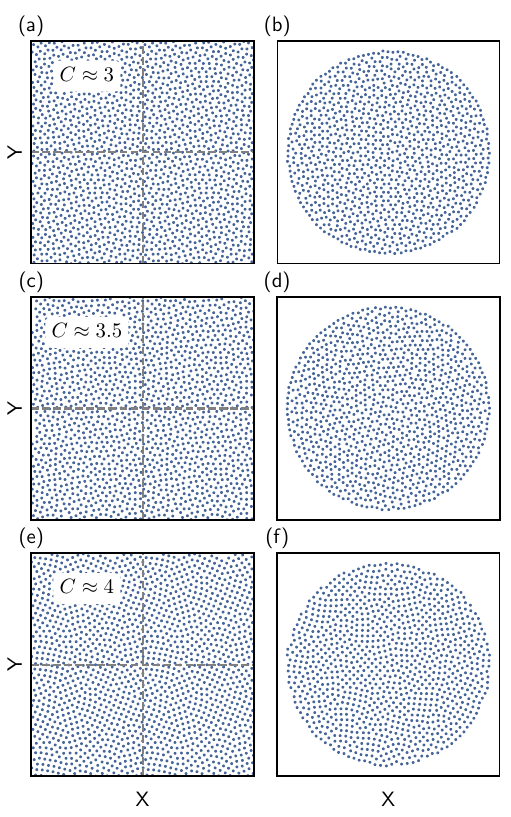}
    \caption{Examples of amorphous solids generated with the variational protocol presented in this article with both periodic (tessellated into a 2x2 grid) and open boundary conditions.}
    \label{fig:examples}
\end{figure}

In Fig.~\ref{fig:examples} we show the entire amorphous solids generated with the variational protocol proposed in this article, each consisting of $N\approx1000$ sites. In particular, we show examples of amorphous solids with $C=3$, $C=3.5$ and $C=4$ with both periodic and open boundaries, exemplifying the range of solids that can be realised with this method.

\section{Linear spin wave theory}\label{app:lswt}
Our starting point is the Ising Hamiltonian introduced in the main text
\begin{equation}\label{eq:ising_hamiltonian_app}
H=-\sum_{i<j}J_{ij}S^z_iS^z_j+h_x\sum_i S^x_i.
\end{equation}
We first minimize the mean-field ground state energy with respect to classical angular variables
\begin{equation}
E_{\text{MF}}= -\frac{1}{4}\sum_{i<j}J_{ij}\cos\theta_j\cos\theta_i-\frac{h_x}{2}\sum_i \sin\theta_i.
\end{equation}
Here we have used a rotation of the initial basis given by 
\begin{equation}
\begin{split}
S^z_i&=\tilde{S}^z_i\cos\theta_i+\tilde{S}^x_i\sin\theta_i, \\
S^x_i&=\tilde{S}^x_i\cos\theta_i-\tilde{S}^z_i\sin\theta_i.
\end{split}
\end{equation}
and the classical assumption that in the transformed basis the state is highly polarized in the $z$ axis, leading to $\langle\tilde{S}^x_i\rangle\simeq 0$. In the new rotated basis, the Hamiltonian reads
\begin{equation}
\begin{split}
\hat{H}=&-\sum_{i<j}J_{ij}\left[\cos\theta_i\cos\theta_j\tilde{S}^z_i\tilde{S}^z_j+\sin\theta_i\sin\theta_j\tilde{S}^x_i\tilde{S}^x_j\right.\\
&\left.+\cos\theta_i\sin\theta_j\tilde{S}^z_i\tilde{S}^x_j+\sin\theta_i\cos\theta_j\tilde{S}^x_i\tilde{S}^z_j\right] \\
&+h_x\sum_i(\cos\theta_i\tilde{S}^x_i-\sin\theta_i\tilde{S}^z_i),
\end{split}
\end{equation}
As the next step, we perform a bosonization of the spins given by:
\begin{equation}
    \begin{split}
        \tilde{S}^z_i = \frac{1}{2}-a\dagg_ia_i,\ \tilde{S}^x_i = \frac{1}{2}(a\dagg_i+a_i).
        \end{split}
\end{equation}
Note that this bosonization is a good approximation only if $\langle a\dagg_ia_i\rangle \ll 1/2$. Under this assumption, we can rewrite $H$ as a quadratic bosonic Hamiltonian 
\begin{equation}
\begin{split}
   H =&-\frac{1}{4}\sum_{i<j} J_{ij} \sin\theta_i\sin\theta_j (a\dagg_i+a_i)(a\dagg_j+a_j)\\
   &+\sum_i \left[h_x\sin\theta_i+\frac{\cos\theta_i}{2}\sum_j J_{ij}\cos\theta_j\right]a\dagg_ia_i. 
\end{split}
    \end{equation}
    Note that here we neglected higher order or constant terms, and the linear term in bosonic operators is proportional to $\frac{\partial E_{\text{MF}}}{\partial \theta_i}=0$. It is useful to rewrite this quadratic Hamiltonian in a compact form:
\begin{equation}
    H =\frac{1}{2}\Psi\dagg M \Psi,
\end{equation}
with the column vector of operators $\Psi \equiv (a_1,\ a_2,\ \dots,\ a_{N}, a\dagg_1,\ a\dagg_2,\ \dots,\ a\dagg_{N})^T$, and the matrix 
\begin{equation}
M\equiv 
2\begin{pmatrix}
G_{ij}+\delta_{ij}H_i & G_{ij} \\
 G_{ij} & G_{ij}+\delta_{ij}H_i 
\end{pmatrix}
\end{equation}
with $G_{ij}=-\frac{1}{8}J_{ij} \sin\theta_i\sin\theta_j$ and $H_i=\frac{1}{2}\left[h_x\sin\theta_i+\frac{\cos\theta_i}{2}\sum_j J_{ij}\cos\theta_j\right]$. Next, we use a Bogoliubov transformation of the bosonic operators to diagonalize the Hamiltonian. We define the Bogoliubov modes via the transformation matrix $U$ such that $\Psi = U\Phi$, with $\Phi\equiv (b_1,\ b_2,\ \dots,\ b_{N}, b\dagg_1,\ b\dagg_2,\ \dots,\ b\dagg_{N})^T$. In terms of the Bogoliubov modes the Hamiltonian reads
\begin{equation}
    H=\frac{1}{2}\Phi\dagg U^\dagger M U\Phi=\frac{1}{2}\Phi\dagg M_D \Phi=
    \frac{1}{2}\Phi\dagg\begin{pmatrix}
\omega_i & 0 \\
0 & \omega_i 
\end{pmatrix} \Phi,
\end{equation}
where $\omega_i$ is the LSWT spectrum. Here the transformation $U$ has the structure 
\begin{equation}
U=\begin{pmatrix}
    u & v \\
    v* & u*
\end{pmatrix}.
    \end{equation}
For systems with open boundary conditions, such as the amorphous solids studied in this work, it is useful to distinguish between bulk and edge Bogoliubov modes. For an amorphous solid centered at the origin and with an open boundary that can be well approximated by a circle of radius $R_\text{edge}$, we define the bulk sites $i$ as the ones satisfying $r_i<R_\text{bulk}$, where $R_\text{bulk}<R_\text{edge}$ is the radius of the subsystem that we consider as the bulk, leading to an edge subsytem consisting of a ring with a width $R_\text{bulk}-R_\text{edge}$. We then consider that a Bogoliubov mode $b_\mu$ is a bulk mode if its contribution from the original bulk modes is larger that a certain threshold $\epsilon_\text{bulk}$, a condition that can be expressed as
\begin{equation}
    \sum_{i\text{ mod }N \in \text{bulk}} |U^{-1}_{\mu, i}\Psi_i |^2<\epsilon_\text{bulk}.
\end{equation}
In this work, we use $\epsilon_\text{bulk}=0.1$.

\section{Simulated annealing: complementary results}\label{app:SA_results}

\subsection{Amorphous solids used in the simulations}
\begin{figure}
  \includegraphics[width=\columnwidth]{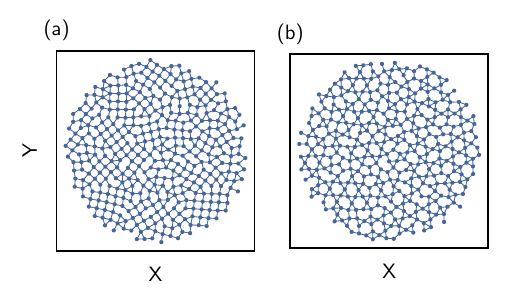}
  \caption{Amorphous solids with $C\approx 4$ used in the SA simulations of the main text. (a) Square-like amorphous solids, with local plaquettes resembling those of a regular square lattice. (b) Kagome-like amorphous solids, with local plaquettes resembling those of a regular kagome lattice.}
  \label{fig:SA_solids}
\end{figure}

In Fig.~\ref{fig:SA_solids} we show the square-like and kagome like solids used in the SA calculations, both with $N=400$ sites.

\subsection{Simulated annealing in fully converged regimes}
\begin{figure}[b]
  \includegraphics{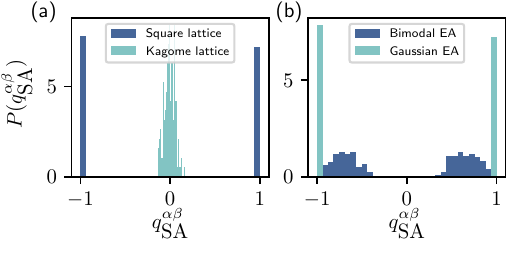}
  \caption{Probability distribution of the replica overlap for well-studied models. (a) Regular square and kagome lattice, with PBC and $N=20\times 20$ and $N=3\times 12\times 12$, respectively. (b) EA models in the square lattice with bimodal and Gaussian couplings, PBC and $N=6\times 6$. In both (a) and (b), we show the results for $n_\text{steps}=10^5$, which are equivalent to those obtained for $n_\text{steps}=10^4$, signalling  SA convergence. }
  \label{fig:simulated_annealing}
\end{figure}

In Fig.~\ref{fig:simulated_annealing} we revisit the properties obtained with the SA method in well-known models to get some intuition. For simplicity, in all these case we use periodic boundary conditions and NN interactions only, and work in regimes of system size and annealing time where SA is able to find the ground state of the system.

Figure~\ref{fig:simulated_annealing}(a) shows the difference between the regular square and kagome lattices with antiferromagnetic NN coupling $-J$, with sizes $N=20\times 20$ and $N=3\times 12\times 12$, respectively. In the square lattice, the presence of a long-range ordered state implies that $P(q_\text{SA}^{\alpha \beta})$ simply consists of two delta functions at $q_\text{SA}^{\alpha \beta}=\pm 1$, as there are only two ground states related by a global spin flip. 
Alternatively, the kagome Ising antiferromagnet model leads to classical spin liquid, and $P(q_\text{SA}^{\alpha \beta})$ exhibits a finite-width peak at $q_\text{SA}^{\alpha \beta}=0$, since the system remains highly disordered even at $T=0$, due to magnetic frustration. For these systems, SA can easily find the ground state even at the large system size under consideration, for moderately slow annealings with $n_\text{steps}>10^3$, due to the simple form of the square lattice antiferromagnet, and the huge degeneracy of the kagome liquid.

In Figure~\ref{fig:simulated_annealing}(b), we consider the EA model on the square lattice, with NN couplings randomly selected from a  bimodal ($\pm J$) distribution, or a Gaussian distribution with zero mean and unit variance.
Here we consider a small system size of $N=6\times 6$, such that we are able to find the exact ground states with SA, despite the complex energy landscape of these systems. For a spin glass state at finite $T$, one of the general features is a nontrivial form of $P(q_\text{SA}^{\alpha \beta})$, as it is directly related to the replica symmetry breaking picture of this phenomenon. In the 2D EA model with bimodal couplings we observe such a nontrivial form of $P(q_\text{SA}^{\alpha \beta})$ at $T=0$, since the model has a degenerate ground state manifold. This also leads to a finite EA parameter $q_\text{SA}^2=0.48$, the order parameter of this spin glass state, which in the thermodynamic limit takes the value $q^2_\text{SA}\approx 0.39$~\cite{thomas_zero-_2011,rubin_dual_2017}. For the Gaussian 2D EA model, the exact ground state is nondegenerate, and therefore $P(q_\text{SA}^{\alpha \beta})$ exhibits a simple two-peak structure at $q_\text{SA}^{\alpha \beta}=\pm 1$, leading to $q_\text{SA}^2=1$.

\begin{figure}[b]
  \includegraphics{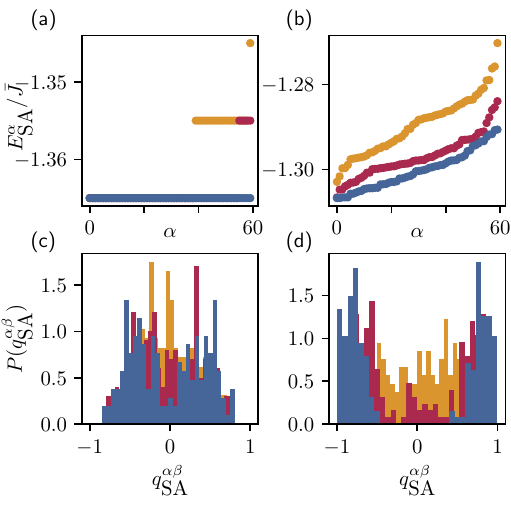}
  \caption{SA results for the bimodal EA model, panels (a) and (c), and a the Gaussian EA model, panels (b) and (d), both with $N=400$. Yellow, purple, and blue colors are used for $n_\text{steps}=10^5,\,10^6,\,$ and $10^7$, respectively. Panels (a) and (b) show the final SA energies of the $N_R=60$ statistically independent replicas. (c) and (d) show the probability distribution of the overlap between the 20 lowest-energy replicas.  }
  \label{fig:simulated_annealing_app}
\end{figure}
\subsection{Simulated annealing of Edwards-Anderson models in the large $N$ case}

For larger system sizes than the ones considered in Fig.~\ref{fig:simulated_annealing}(b), the EA models typically require large annealing times in order to find the ground state of the system, due to the complex low-energy landscape of these models, which are known to host a spin glass state at $T=0$~\cite{kawashima_critical_1992, campbell_energy_2004,houdayer_low-temperature_2004,rieger_critical_1996,rubin_dual_2017,xu_dynamic_2017}. To illustrate this behavior, in Fig.~\ref{fig:simulated_annealing_app} we show the SA results for both the bimodal and Gaussian EA models, for $N=400$. In this case, we observe that the replica energies, Figs.~\ref{fig:simulated_annealing_app}(a)-(b), still depend on the annealing time for $n_\text{steps}>10^6$. We also observe differences between the bimodal and Gaussian cases. The bimodal EA model seems to have converged, at least for the lowest energy replicas, for the largest SA run with $n_\text{steps}=10^7$. This is because such a model has an infinite ground state degeneracy in the thermodynamic limit, which makes it easier for the SA to find one of such solutions. In contrast, we observe that the Gaussian EA model converges much slower, since this model has a nondegenerate ground state (up to a global spin flip).  Indeed, without annealed solutions corresponding to excited states, the Gaussian EA model should exhibit the same two delta functions as in Fig.~\ref{fig:simulated_annealing}(b).

\bibliographystyle{apsrev4-2}

\end{document}